\pdfoutput=1

\documentclass{article}
\usepackage{spconf,amsmath,graphicx}

\usepackage{tabu}                           % Better tables
\usepackage{multirow,makecell}              % Cells spanning multiple rows in tables

\usepackage{xspace}
\def\etal{\textit{et al.}\xspace}

\usepackage[bibstyle=ieee,sorting=none,giveninits=true,citestyle=numeric]{biblatex}
\AtNextBibliography{\footnotesize}

\usepackage[pdftex,colorlinks]{hyperref}
\usepackage[all]{hypcap}    % Make links jump to the start of figures, not the caption

\title{\vspace{-1.5em}A COMPARISON OF FIVE MULTIPLE INSTANCE LEARNING POOLING FUNCTIONS
FOR SOUND EVENT DETECTION WITH WEAK LABELING}

\name{Yun Wang$^\dag$, Juncheng Li\,$^{\dag,\ddag}$, and Florian Metze$^\dag$
\thanks{\scriptsize The first two authors were supported by
a graduate research fellowship award from Robert Bosch LLC;
the first author was also supported by a faculty research award from Google.
This work used the ``comet'' and ``bridges'' clusters of the XSEDE environment \cite{XSEDE},
supported by NSF grant number ACI-1548562.}}
\address{$^\dag$Language Technologies Institute, Carnegie Mellon University, Pittsburgh, PA, U.S.A. \\
$^\ddag$Research and Technology Center, Robert Bosch LLC, Pittsburgh, PA, U.S.A. \\
\texttt{maigoakisame@gmail.com, \{junchenl, fmetze\}@cs.cmu.edu}}

\begin{document}
\ninept

\maketitle

\begin{abstract}
Sound event detection (SED) entails two subtasks: recognizing what types of
sound events are present in an audio stream (audio tagging),
and pinpointing their onset and offset times (localization).
In the popular multiple instance learning (MIL) framework
for SED with weak labeling, an important component is the pooling function.
This paper compares five types of pooling functions
both theoretically and experimentally,
with special focus on their performance of localization.
Although the attention pooling function is currently receiving the most attention,
we find the linear softmax pooling function to perform the best among the five.
Using this pooling function, we build a neural network called TALNet.
It is the first system to reach state-of-the-art audio tagging performance
on Audio Set, while exhibiting strong localization performance
on the DCASE 2017 challenge at the same time.
\end{abstract}


\section{Introduction}
\label{sec:intro}

Sound event detection (SED) is the task of detecting the type,
onset time and offset time of sound events in audio stream.
While some studies are satisfied with recognizing what types
of sound events are present in a recording (\emph{audio tagging}),
this paper pays special attention to the \emph{localization}
of sound events.

Modern SED systems usually take the form of neural networks,
with convolutional layers~\cite{gorin2016dcase, espi2015exploiting},
recurrent layers~\cite{Yun-ICASSP2016, parascandolo2016recurrent, adavanne2016sound, hayashi2016bidirectional},
or both~\cite{cakir2017convolutional}.
The networks predict the probability of each sound event type
frame by frame; applying a threshold to these frame-level probabilities
will then produce localized detections of sound events.

Traditionally, the training of SED models relied upon \emph{strong labeling},
which specifies the type, onset time and offset time of each sound event occurrence.
But such annotation is very tedious to obtain by hand.
In order to scale SED up, researchers have turned to SED with \emph{weak labeling},
which only specifies the types of sound events present in each training recording
but does not provide any temporal information.
In March 2017, Google released the weakly labeled Audio Set~\cite{AudioSet},
which is by far the largest corpus available for SED.
The DCASE challenge of 2017~\cite{DCASE2017} featured a task of SED
with weak labeling, which used a subset of Audio Set.

\begin{figure}[t]
\centering
\includegraphics[width=0.83\columnwidth]{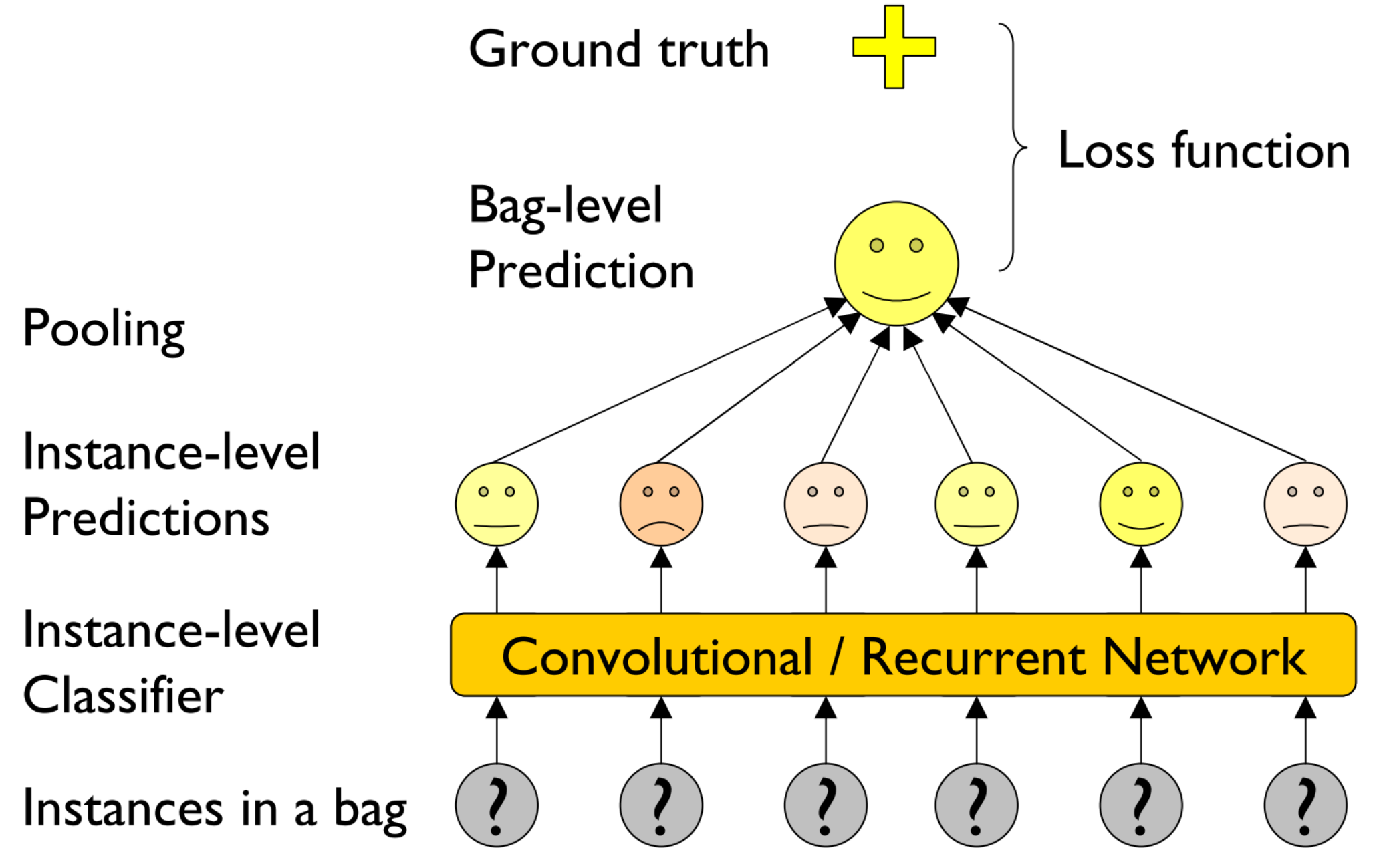}
\caption{Block diagram of a MIL system for SED with weak labeling.}
\label{fig:mil-for-sed}
\end{figure}

A common framework for SED with weak labeling
is \emph{multiple instance learning} (MIL)~\cite{amores2013multiple},
as shown in Fig.~\ref{fig:mil-for-sed}.
In MIL, we do not know the ground truth label of every training instance;
instead, the instances are grouped into bags,
and we only know the label of bags.
In the case of binary classification,
the relationship between instance labels and bag labels
often obey the \emph{standard multiple instance} (SMI) \emph{assumption}:
the bag label is positive if and only if the bag contains at least
one positive instance.
In SED, each training recording is regarded as a bag,
and its frames are regarded as instances.
Each sound event type is considered independently,
so SED becomes a binary classification problem for each sound event type.
A neural network predicts the probability of each sound event type
being active at each frame.
Then, a \emph{pooling function} aggregates the frame-level probabilities
into a recording-level probability for each sound event type.
The recording-level probabilities can be compared against the recording labels
to compute a loss function,
and the network can then be trained to minimize the loss.

The choice of the pooling function is an important decision.
The default choice is the ``max'' pooling function~\cite{su2017weakly, kumar2016audio},
which is faithful to the SMI assumption.
A previous study of ours~\cite{Yun-Interspeech2018}
has evaluated a ``noisy-or'' pooling function%
~\cite{maron1998framework, zhang2006multiple, babenko2008simultaneous},
and has shown it does not work for localization
despite its nice probabilistic interpretation under the SMI assumption.
Since the 2017 DCASE challenge, a number of other pooling functions
have been reported to perform well even though they deviate from the SMI assumption.
These include average pooling~\cite{shah2018closer},
two softmax pooling functions based on linear weighting~\cite{dang2017deep}
and exponential weighting~\cite{salamon2017dcase},
as well as an attention-based pooling function~\cite{xu2018large, kong2018audio}.
The purpose of this paper is to compare these pooling functions
against max pooling from two aspects:
theoretically, we derive the gradient of the five pooling functions,
and check if their signs lead the training down the right way;
experimentally, we compare the five pooling functions on two SED corpora:
the DCASE 2017 challenge~\cite{DCASE2017} and Audio Set~\cite{AudioSet}.
Although the attention pooling function appears to be the most favored by researchers,
we demonstrate that it is the linear softmax pooling function
that works best for localization.

Our experiments also result in a convolutional and recurrent neural network (CRNN)
which is the first system within our knowledge that
exhibits strong performance on audio tagging and localization at the same time.
We name this network ``TALNet'', where ``TAL'' stands for
``tagging and localization''.
This network closely matches the current state-of-the-art
audio tagging performance on Audio Set,
while achieving competitive localization performance
on the DCASE 2017 challenge without any finetuning.

\section{Theoretical Comparison of the Five Pooling Functions}
\label{sec:theory}

\subsection{Definition of the Pooling Functions}
\label{sec:definition}

Let $y_i \in [0,1]$ be the predicted probability of a certain event type at the $i$-th frame,
and $y \in [0,1]$ be the aggregated recording-level probability of the same event.
We list the definitions of the five pooling functions to be compared
in Table~\ref{table:pool}.

The max pooling function simply takes the largest $y_i$ to be $y$.
If the same threshold is applied to the recording-level and
frame-level probabilities,
then the frame-level predictions and recording-level prediction
are guaranteed to be consistent with the SMI assumption.
However, the max pooling function has a defect that
only one frame in a recording can receive an error signal.
As a consequence, if an event occurs multiple times in a recording,
the occurrences that do not cover this frame may be easily missed.
All the other four pooling functions
try to alleviate this problem by assigning some weight
to smaller $y_i$'s when aggregating them to produce $y$.

The average pooling function~\cite{shah2018closer}
assigns an equal weight to all frames.
The equation appears to defy the SMI assumption,
but it is reported to perform better than the max pooling function
in~\cite{shah2018closer}.

The two softmax pooling functions compute $y$
as a weighted average of the $y_i$'s,
where larger $y_i$'s receive larger weights.
In this way, the recording-level probability is still mainly
determined by the larger frame-level probabilities,
but frames with smaller probabilities get a chance
to receive an error signal.
The linear softmax function~\cite{dang2017deep}
assigns weights equal to the frame-level probabilities $y_i$ themselves,
while the exponential softmax function~\cite{salamon2017dcase}
assigns a weight of $\exp(y_i)$ to the frame-level probability $y_i$.

Finally, in the attention pooling function~\cite{xu2018large, kong2018audio},
the weights for each frame $w_i$ are learned
with a dedicated layer in the network.
The recording-level probability $y$ is then computed using the general
weighted average formula.
The attention pooling function appears to be most favored by researchers
because of its flexibility, and variants have emerged such as
the multi-level attention in~\cite{yu2018multi}.

\subsection{Gradient of the Pooling Functions}
\label{sec:gradient}

In this section, we analyze the gradient of the loss function
w.r.t. the frame-level probabilities $y_i$
(and, in the case of attention, also the weights $w_i$).
Let $t \in \{0,1\}$ be the recording-level ground truth.
The loss function is usually the cross entropy:
\begin{equation}
L = -t \log y - (1-t) \log (1-y)
\end{equation}
We decompose its gradient with respect to the frame-level probabilities $y_i$
(and the frame-level weights $w_i$) using the chain rule:
\begin{equation}
\frac{\partial L}{\partial y_i} = \frac{\partial L}{\partial y} \frac{\partial y}{\partial y_i}, \quad \quad
\frac{\partial L}{\partial w_i} = \frac{\partial L}{\partial y} \frac{\partial y}{\partial w_i}
\label{eq:chain}
\end{equation}
The first term,
\begin{equation}
\frac{\partial L}{\partial y} = -\frac{t}{y} + \frac{1-t}{1-y}
\end{equation}
does not depend on the choice of the pooling function.
It is negative when the recording label is positive ($t = 1$),
and positive when the recording label is negative ($t = 0$).
The second term, $\partial y / \partial y_i$ (and $\partial y / \partial w_i$),
is calculated for each pooling function in Table~\ref{table:pool}.

\begin{table}[t]
\centering
\tabulinesep=0.2em
\resizebox{\columnwidth}{!}{
\begin{tabu}{r|l|l}
\hline
& \bf{Pooling Function} & \bf{Gradient} \\
\hline
\bf{Max pooling} & $\displaystyle y = \max_i y_i$ & $\displaystyle \frac{\partial y}{\partial y_i} = \begin{cases} 1, & \text{if } y_i = y \\ 0, & \text{otherwise} \end{cases}$ \\
\hline
\bf{Average pooling} & $\displaystyle y = \frac{1}{n} \textstyle \sum_i y_i$ & $\displaystyle \frac{\partial y}{\partial y_i} = \frac{1}{n}$ \\
\hline
\bf{Linear softmax} & $\displaystyle y = \frac{\sum_i y_i^2}{\sum_i y_i}$ & $\displaystyle \frac{\partial y}{\partial y_i} = \frac{2y_i - y}{\sum_j y_j}$ \\
\hline
\bf{Exp. softmax} & $\displaystyle y = \frac{\sum_i y_i \exp(y_i)}{\sum_i \exp(y_i)}$ & $\displaystyle \frac{\partial y}{\partial y_i} = (1 - y + y_i) \cdot \frac{\exp(y_i)}{\sum_j \exp(y_j)}$ \\
\hline
\bf{Attention} & $\displaystyle y = \frac{\sum_i y_i w_i}{\sum_i w_i}$ & $\displaystyle \frac{\partial y}{\partial y_i} = \frac{w_i}{\sum_j w_j}, \quad \frac{\partial y}{\partial w_i} = \frac{y_i - y}{\sum_j w_j}$ \\
\hline
\end{tabu}
}
\caption{The five pooling functions and their gradients.
$n$ is the number of frames in a recording.}
\label{table:pool}
\end{table}

With the max pooling function, $\partial y / \partial y_i$
equals 1 for the frame with the largest probability and 0 elsewhere.
The fact that only one frame receives a non-zero gradient
may cause many frame-level false negatives.
The gradient for this single frame, though, does have the correct sign:
when $t = 1$, the gradient $\partial L / \partial y_i$ is negative,
so the frame-level probability $y_i$ will be boosted in order to reduce the loss;
when $t = 0$, the gradient is positive, so $y_i$
will be suppressed.

With the average pooling function, $\partial y / \partial y_i$
equals $1/n$ regardless of the value of $y_i$.
This means the gradient is distributed evenly across all frames.
For negative recordings, this will suppress the probability $y_i$ of all frames,
and this is correct behavior.
For positive recordings, however, not all frames should be boosted,
and the average pooling function can produce a lot of false positive frames.

With the linear softmax pooling function,
$\partial y / \partial y_i$ is positive where $y_i > y/2$,
which gives rise to complicated and interesting behavior.
For positive recordings ($t = 1$), the gradient
is negative where $y_i > y/2$, and positive where $y_i < y/2$.
As a result, larger $y_i$'s will be boosted, while smaller $y_i$'s will be suppressed.
This is exactly the desired behavior under the SMI assumption:
the frame-level probabilities are driven to the extremes 0 and 1,
resulting in well-localized detections of sound events.
For negative recordings ($t = 0$), the gradient
is positive where $y_i > y/2$, and negative where $y_i < y/2$.
This means all frame-level probabilities will be pushed toward $y/2$.
Considering that $y$ is a weighted average of the $y_i$'s,
given enough iterations, all the $y_i$'s will converge to zero as desired.

With the exponential pooling function, $\partial y / \partial y_i$
is always positive, just like with the average pooling function.
As a result, the exponential pooling function also has the concern of
producing too many false positive frames.
Nevertheless, the problem will be less serious,
because smaller $y_i$'s receive smaller gradients.

\begin{figure*}[t]
\centering
\includegraphics[angle=90,height=8.7em]{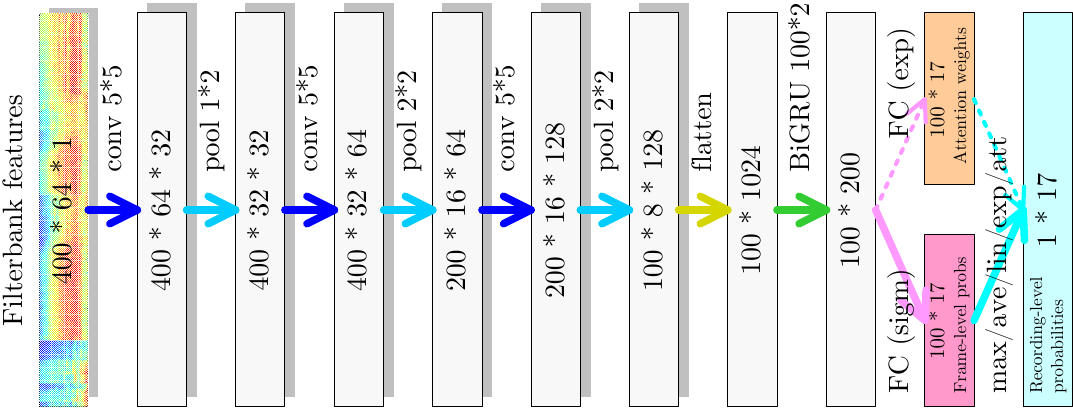}
\hfill
\includegraphics[angle=90,height=8.7em]{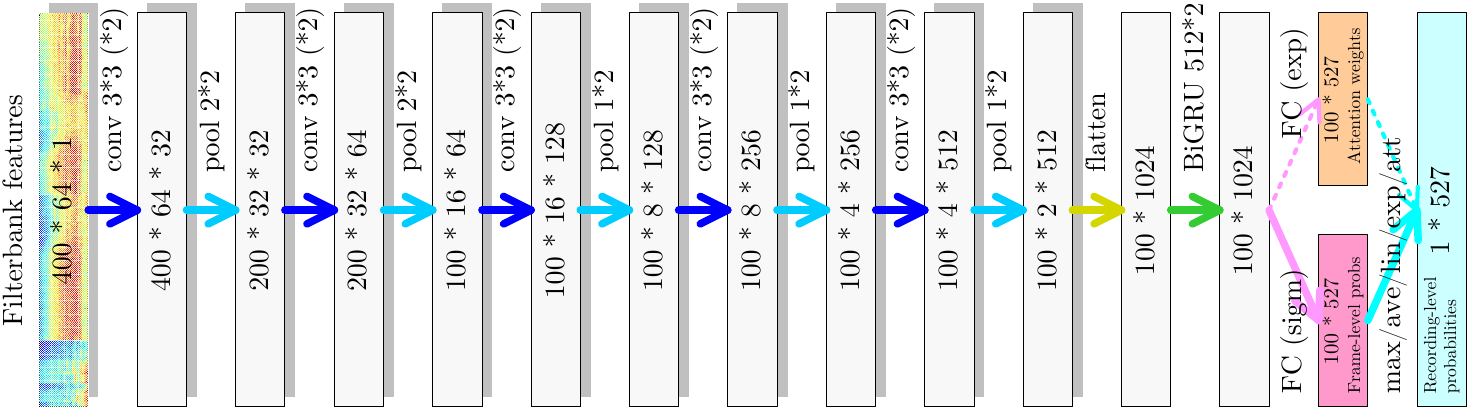}
\caption{Structures of the networks used in
Secs.~\ref{sec:dcase} (left) and \ref{sec:talnet} (right).
The shape is specified as ``frames * frequency bins * feature maps''
for \mbox{3-D} tensors (shaded), and
``frames * feature maps'' for \mbox{2-D} tensors.
``conv $n$*$m$'' stands for a convolutional layer with the specified
kernel size and ReLU activation; ``(*2)'' means
the layer is repeated twice.
``pool $n$*$m$'' stands for a max pooling layer with the specified stride.
``FC'' is short for ``fully connected''.
At the output end, the ``attention weights'' block is only used
with the attention pooling function.\vspace{-0.5em}}
\label{fig:structures}
\end{figure*}

With the attention pooling function, the term
$\partial y / \partial y_i$ is always positive.
Therefore, the frame-level probabilities will be boosted or suppressed
according to the recording label, with strengths proportional
to the learned weights.
This is correct behavior if frames with larger probabilities $y_i$
also get larger weights $w_i$.
However, because the weights $w_i$ are also learned,
we should also consider $\partial y / \partial w_i$,
the gradient of the loss function w.r.t. the weights:
this term is positive where $y_i > y$.
When the recording is positive, this will cause the weight $w_i$
to rise where the frame-level probability $y_i$ is large
and to shrink where the $y_i$ is small,
agreeing with the motivation that frames with larger probabilities $y_i$
should get larger weights $w_i$.
When the recording is negative, however, the opposite phenomenon will happen:
larger weights will concentrate upon frames with smaller probabilities.
This has a serious consequence:
while the recording-level probability $y$ will indeed be small,
there will be frames with large probabilities $y_i$ and small weights $w_i$.
This means the recording-level prediction and frame-level predictions
will be inconsistent with the SMI assumption,
and the frames with large probabilities will end up being false positives
for localization.

\section{Experimental Comparison of the Five Pooling Functions}
\label{sec:exp}

\subsection{The DCASE 2017 Challenge}
\label{sec:dcase}

We first compared the five pooling functions on Task~4 of the
DCASE 2017 challenge~\cite{DCASE2017}.
The task involves 17~types of vehicle and warning sounds,
and evaluates both tagging and localization.

The data used in the task is a subset of Audio Set~\cite{AudioSet}.
It consists of a training set (51,172~recordings),
a public test set (488~recordings),
and a private evaluation set (1,103~recordings).
All the recordings are 10-second excerpts from YouTube videos.
The test and evaluation sets are strongly labeled so
they can be used to evaluate both audio tagging and localization,
but the training set only comes with weak labeling.
Because we did not have access to the ground truth of the evaluation set,
we report the performance of our systems on the test set.
The test and evaluation sets have balanced numbers of the events,
but the training set is unbalanced.
We set aside 1,142~recordings from the training set
to make a balanced validation set, and used the remaining 50,030~recordings for training.

We implemented a convolutional and recurrent neural network (CRNN),
whose structure is shown in Fig.~\ref{fig:structures} (left).
The input is a matrix of filterbank features;
it has 400~frames and 64~frequency bins.
The convolutional and pooling layers reduce the frame rate from 40~Hz to 10~Hz.
At the output end, a fully connected layer with sigmoid activation
produces frame-level predictions, which are then aggregated across time into
recording-level predictions using any of the five pooling functions.
If the attention pooling function is used, a separate fully connected layer
with exponential activation is used to generate the weights.
The network was trained using the PyTorch toolkit~\cite{PyTorch}.
We applied data balancing so each minibatch
contained roughly equal numbers of recordings of each event type.

The performance of audio tagging was evaluated with the
micro-average $F_1$ on the recording level;
localization was evaluated with the
micro-average error rate (ER) and $F_1$ on 1-second segments.
The $F_1$ is the larger the better while the error rate is the smaller the better;
refer to~\cite{sed-eval} for detailed definitions of these evaluation metrics.
To make binary predictions on the recording level,
we thresholded the recording-level probabilities with class-specific thresholds,
which were tuned to optimize the audio tagging $F_1$ on the validation set.
To make binary predictions on 1-second segments,
we first computed segment-level probabilities
by aggregating the 10~frame-level probabilities within each segment,
then thresholded the segment-level probabilities
using the same class-specific thresholds.

\begin{table}[t]
\centering
\resizebox{\columnwidth}{!}{
\begin{tabular}{r|c|c|c|c|c}
\hline
& \bf{Max Pool.} & \bf{Ave. Pool.} & \bf{Lin. Soft.} & \bf{Exp. Soft.} & \bf{Attention} \\
\hline
\multicolumn{1}{l|}{\bf{Audio Tagging} \quad} & & & & & \\
\bf{TP}         &  284 &  297 &  317 &  298 &  301 \\
\bf{FN}         &  322 &  309 &  289 &  308 &  305 \\
\bf{FP}         &  364 &  285 &  359 &  324 &  317 \\
\bf{Precision}  & 43.8 & 51.0 & 46.9 & 47.9 & 48.7 \\
\bf{Recall}     & 46.9 & 49.0 & 52.3 & 49.2 & 49.7 \\
\bf{$F_1$}      & \bf{45.3} & \bf{50.0} & \bf{49.5} & \bf{48.5} & \bf{49.2} \\
\hline
\multicolumn{1}{l|}{\bf{Localization} \quad} & & & & & \\
\bf{TP}         & 1,206 & 2,114 & 1,832 & 2,121 & 1,926 \\
\bf{FN}         & 3,154 & 2,246 & 2,528 & 2,239 & 2,434 \\
\bf{FP}         & 1,253 & 3,758 & 2,187 & 3,437 & 3,309 \\
\bf{Precision}  &  49.0 &  36.0 &  45.6 &  38.2 &  36.8 \\
\bf{Recall}     &  27.7 &  48.5 &  42.0 &  48.6 &  44.2 \\
\bf{$F_1$}      & \bf{35.4} & \bf{41.3} & \bf{43.7} & \bf{42.8} & \bf{40.1} \\
\hline
\multicolumn{1}{l|}{\bf{Localization} \quad} & & & & & \\
\bf{Sub.}       &   712 & 1,385 & 1,040 & 1,292 & 1,275 \\
\bf{Del.}       & 2,442 &   861 & 1,488 &   947 & 1,159 \\
\bf{Ins.}       &   541 & 2,373 & 1,147 & 2,145 & 2,034 \\
\bf{Error Rate} & \bf{84.7} & \bf{105.9} & \bf{84.3} & \bf{100.6} & \bf{102.5} \\
\hline
\end{tabular}
}
\caption{Detailed performance of the five systems on Task~4 of the DCASE 2017 challenge.
Error rates and $F_1$'s are in percentages.}
\label{table:dcase-performance}
\end{table}

Table~\ref{table:dcase-performance} compares the performance of the five pooling functions.
All the four new pooling functions outperform max pooling
in terms of $F_1$ for both audio tagging and localization.
In terms of localization error rate, however, only the linear softmax system
slightly outperforms max pooling;
the average pooling, exponential softmax and attention systems
yield error rates over 100\%.

Table~\ref{table:dcase-performance} also includes a breakdown of the error types.
All the five pooling functions achieve a reasonable balance between
false negatives and false positives for audio tagging.
However, the breakdown of errors for localization reveals that
only the linear softmax pooling function maintains a good balance.
As analyzed in Sec.~\ref{sec:gradient},
the max pooling system makes too many false negatives,
which result in a low recall and a low $F_1$;
the average, exponential softmax and attention pooling functions
make too many false positives, which result in a high insertion rate
and consequently a high error rate.

\begin{figure}[t]
\centering
\includegraphics[width=0.95\columnwidth]{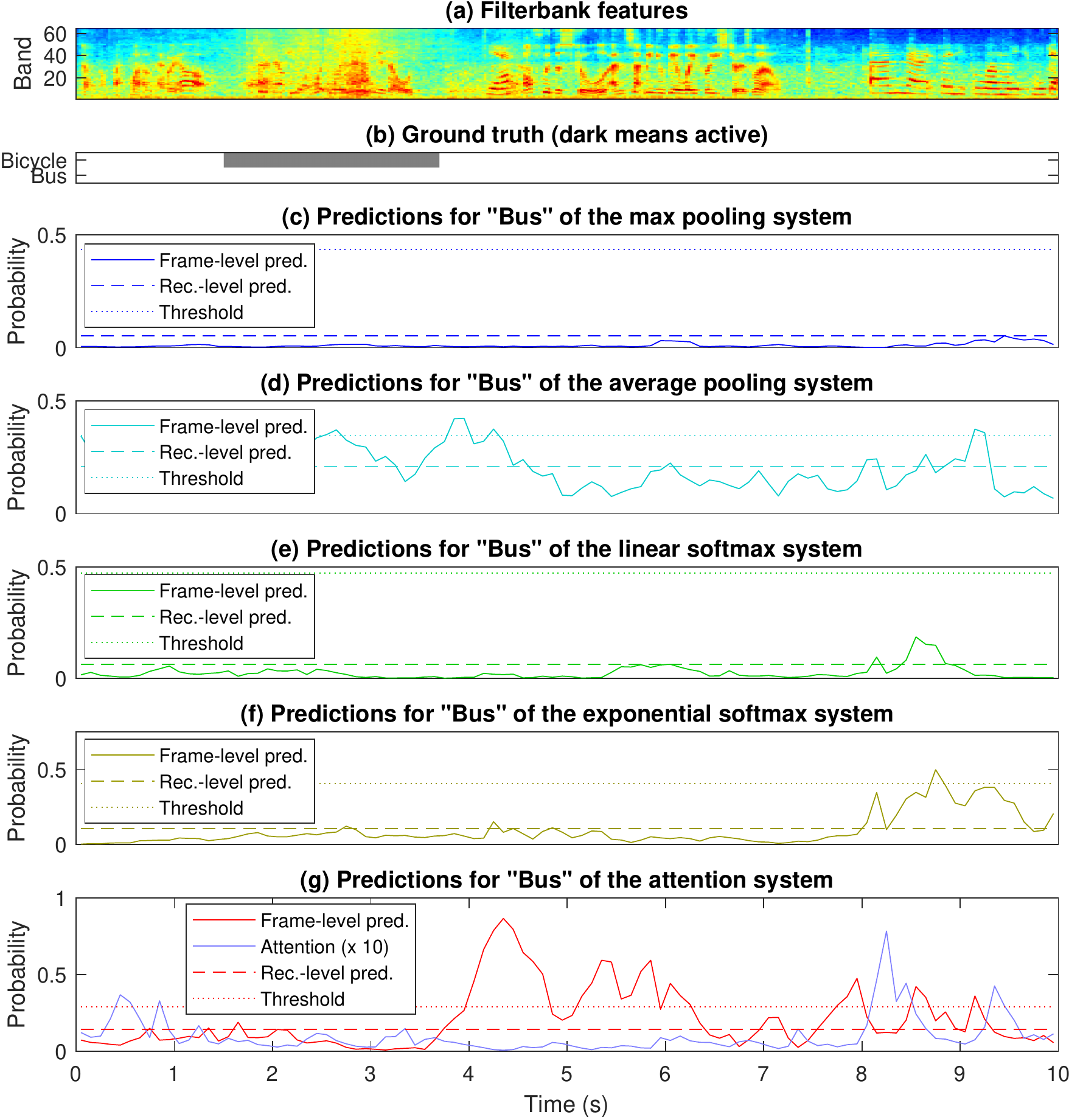}
\vspace{-1em}
\caption{The frame-level predictions of the five systems
for the \texttt{bus} event on the test recording ``\texttt{-nqm\char`_RJ2xj8}''
(unfortunately, this recording is no longer available on YouTube).
Best viewed in color.}
\label{fig:dcase-example-2}
\end{figure}

Fig.~\ref{fig:dcase-example-2} illustrates the false positives made by
the average pooling, exponential softmax and attention systems.
The recording contains speech and bicycle noise, but no bus noise.
On the recording level, all five systems correctly detect the bicycle sound
and deny the existence of bus sounds.
On the frame level, the max and linear softmax systems predict low probabilities
that safely stay below the threshold for \texttt{bus} throughout the recording.
In the average pooling and exponential softmax systems, however,
some frame-level probabilities exceed the threshold,
even though other frames with lower probabilities keep
the recording-level probability under control.
In the attention system, we see exactly what we have anticipated in Sec.~\ref{sec:gradient}:
the attention (light blue line) mostly focuses on regions
where the frame-level probabilities are low (8.2s).
This correctly produces a negative recording-level prediction,
but lets many frame-level false positives (4$\sim$6s) get away unconstrained.
The false positives shown in Fig.~\ref{fig:dcase-example-2}
are common throughout the data.

Considering the balance between false negatives and false positives for localization,
as well as the agreement between recording-level and frame-level predictions,
we recommend the linear softmax pooling function among all the pooling functions
we have studied.

\subsection{TALNet: Joint Tagging and Localization on Audio Set}
\label{sec:talnet}

We also compared the five pooling functions on the entire Audio Set~\cite{AudioSet}.
This corpus provides a training set of over 2~million recordings,
and a evaluation set of 20,371 recordings.
The recordings in both sets are 10-second YouTube video excerpts,
labeled with the presence or absence of 527~types of sound events.

We trained a CRNN with the structure shown in Fig.~\ref{fig:structures} (right).
We name the network ``TALNet'', where ``TAL'' stands for
``tagging and localization''.
The network has 10~convolutional layers, 5~pooling layers, and 1~recurrent layer.
We applied data balancing during training;
we also found it essential to apply batch normalization~\cite{batchnorm}
before the ReLU activation of each convolutional layer.

Audio Set only provides evaluation metrics for audio tagging.
These include the mean average precision (MAP),
mean area under the curve (MAUC), and d-prime ($d'$);
all these metrics are the larger the better.
To evaluate the performance of localization,
we applied TALNet to the DCASE 2017 challenge directly,
and measured the same metrics as in Sec.~\ref{sec:dcase}.
The results are listed in the top row of Table~\ref{table:talnet-performance}.
Although the linear softmax system is not the best
in terms of all the evaluation metrics,
it is the only system that achieves a low error rate
and a high $F_1$ for localization.
The max pooling system falls behind on the $F_1$,
while the other three systems exhibit excessively high error rates.

\begin{table}[t]
\centering
\resizebox{\columnwidth}{!}{
\setlength{\tabcolsep}{0.55em}
\begin{tabular}{c|c|c|c|c|c|c|c}
\hline
\multirowcell{2}{\bf{System}} & \multirowcell{2}{\bf{\begin{tabular}{@{}c@{}}\# Train \\ Recs.\end{tabular}}} & \multicolumn{3}{c|}{\bf{Audio Set}} & \multicolumn{3}{c}{\bf{DCASE 2017}} \\
\cline{3-8}
& & \bf{MAP} & \bf{MAUC} & \bf{d'} & \bf{Tag.F1} & \bf{Loc.ER} & \bf{Loc.F1} \\
\hline
Max pooling     & \multirowcell{5}{2M} & 0.351 & 0.961 & 2.497 & 52.6 & 81.5 & 42.2 \\
Average pooling & & 0.361 & \bf{0.966} & 2.574 & \bf{53.8} & 101.8 & \bf{46.8} \\
Linear softmax  & & 0.359 & \bf{0.966} & \bf{2.575} & 52.3 & \bf{78.9} & 45.4 \\
Exp. softmax    & & \bf{0.362} & 0.965 & 2.554 & 52.3 & 89.2 & 46.2 \\
Attention       & & 0.354 & 0.963 & 2.531 & 51.4 & 92.0 & 45.5 \\
\hline
Hershey~\cite{hershey2017cnn, AudioSet} &   1M & 0.314 & 0.959 & 2.452 & & & \\
Kumar~\cite{kumar2017knowledge}         &  22k & 0.213 & 0.927 & & & & \\
Shah~\cite{shah2018closer}              &  22k & 0.229 & 0.927 & & & & \\
Wu~\cite{wu2017reducing}                &  22k &       & 0.927 & & & & \\
Kong~\cite{kong2018audio}               &   2M & 0.327 & 0.965 & 2.558 & & & \\
Yu~\cite{yu2018multi}                   &   2M & \bf{0.360} & \bf{0.970} & \bf{2.660} & & & \\
Chen~\cite{chen2018class}               & 600k & 0.316 & & & & & \\
Chou~\cite{chou2018learning}            &   1M & 0.327 & 0.951 & & & & \\
\hline
\end{tabular}
}
\caption{The performance of TALNet on both Audio Set and the DCASE 2017 challenge,
compared with various systems in the literature (not all used the full training set).
Bold font indicates the best performance in each group.}
\label{table:talnet-performance}
\end{table}

In Table~\ref{table:talnet-performance} we also
list the results on Audio Set reported in all the literature that we can find.
Our system closely matches the system of Yu \etal~\cite{yu2018multi},
and outperforms all other systems by a large margin.
We would like to point out that the systems in the literature
either do not perform localization well, or do not perform localization at all.
For example, the system of Kong \etal~\cite{kong2018audio}
uses the attention pooling function.
As we have demonstrated, this pooling function can cause
many false positives on the frame level, and suffer from a high error rate.
The system of Yu \etal~\cite{yu2018multi} uses multi-level attention:
attention layers are built upon multiple hidden layers,
whose outputs are concatenated and further processed by a fully connected layer
to yield a recording-level prediction.
No frame-level predictions at all are made in this process.
In contrast, our TALNet is the first system we know that
achieves good performance for both audio tagging and localization at the same time.

\section{Conclusion and Discussion}
\label{sec:conclusion}

In this paper we have compared five pooling functions,
and shown linear softmax to be the best among the five.
The linear softmax pooling function has the following advantages:
(1) it allows the gradient to flow unobstructedly;
(2) it achieves a balance between false negatives and false positives for localization;
(3) its predictions on the recording level and the frame level are relatively consistent.
Using the linear softmax pooling function, we have built TALNet,
which is the first network to achieve a strong performance
for both audio tagging and localization at the same time.
Our findings may not be limited to SED, but can apply generally to any MIL problem.

Nevertheless, linear softmax is by no means the ultimate optimal pooling function.
An adaptive pooling function has been proposed in~\cite{mcfee2018adaptive};
it gives a weight of $\exp(\alpha y_i)$ to the frame-probability $y_i$,
and may be considered a generalization of the exponential softmax pooling function.
Along this line, we may also consider a weighting scheme of $y_i^\beta \exp(\alpha y_i)$,
which would subsume both the linear softmax and the exponential softmax poling functions.

At the same time, the flexibility of the attention pooling function
to learn the weights on the fly is still attractive,
despite its excessive false positives on the frame level.
We have found that the false positives are caused by
the attention focusing on frames with low probabilities.
We believe the attention pooling function is still promising
if we could somehow impose a constraint of monotonicity
that frames with larger probabilities must receive larger weights.

For more details about the experiments, such as the data balancing algorithm,
how the class-specific thresholds were tuned,
and the hyperparameters for training, please refer to Chapter~3
of the first author's PhD thesis~\cite{Yun-PhD-Thesis}.
The code and acoustic features for the experiments are available at
{\footnotesize \url{https://github.com/MaigoAkisame/cmu-thesis}}.

\vfill\pagebreak

\section{REFERENCES}
\label{sec:refs}

\printbibliography[heading=none]

\end{document}